\documentclass[twocolumn,showpacs,preprintnumbers,amsmath,amssymb,prx]{revtex4}
\usepackage{tikz}
\usetikzlibrary{calc,decorations.markings}
\usepackage{epsfig}
\usepackage{graphicx}
\usepackage{rotating}
\usepackage{amssymb}
\usepackage{amsmath}
\usepackage{float}   
\usepackage[active]{srcltx}
\usepackage{url}
\usepackage{hyperref}
\usepackage{centernot}
\usepackage{dsfont,bm}
\usepackage{xcolor}
\usepackage{pgfplots}
\hypersetup{
	colorlinks   = true, 
	urlcolor     = blue, 
	linkcolor    = blue, 
	citecolor   = blue 
}

\DeclareMathOperator{\tr}{tr}

\usepackage{amsthm}
\newtheorem{prop}{Proposition}
\newtheorem{lem}{Lemma}

\graphicspath{{figs/}}

\begin{document}

\title{Quantum Trajectory Approach to Error Mitigation}

\author{Brecht I. C Donvil}
\email{brecht.donvil@uni-ulm.de}
\affiliation{Institute for Complex Quantum Systems and IQST, Ulm University - Albert-Einstein-Allee 11, D-89069 Ulm, Germany}

\author{Rochus Lechler}
\affiliation{Institute for Complex Quantum Systems and IQST, Ulm University - Albert-Einstein-Allee 11, D-89069 Ulm, Germany}

\author{Paolo Muratore-Ginanneschi}
\affiliation{University of Helsinki, Department of Mathematics and Statistics, P.O. Box 68 FIN-00014, Helsinki, Finland}

\author{Joachim Ankerhold}
\affiliation{Institute for Complex Quantum Systems and IQST, Ulm University - Albert-Einstein-Allee 11, D-89069 Ulm, Germany}

\begin{abstract}
Quantum Error Mitigation (EM) is a collection of strategies to reduce errors on noisy intermediate scale quantum (NISQ) devices on which proper quantum error correction is not feasible. One of such strategies aimed at mitigating noise effects of a known environment is to realise the inverse map of the noise using a set of completely positive maps weighted by a quasi-probability distribution, i.e. a probability distribution with positive and negative values. This quasi-probability distribution is realised using classical post-processing after final measurements of the desired observables have been made. Here we make a connection with quasi-probability EM and recent results from quantum trajectory theory for open quantum systems. We show that the inverse of noise maps can be realised by performing classical post-processing on the quantum trajectories generated by an additional reservoir with a quasi-probability measure called the influence martingale. We demonstrate our result on a model relevant for current NISQ devices. Finally, we show the quantum trajectories required for error correction can themselves be simulated by coupling an ancillary qubit to the system.
\end{abstract}
\maketitle


\section{Introduction}
Current quantum computation platforms operate in the noisy intermediate scale quantum (NISQ) regime. The noisy character of these devices significantly inhibits their ability to successfully perform quantum computations. Managing and reducing this noise is therefore of the main challenges in current era quantum platforms.

The main strategy to counteract noise is quantum error correction, see e.g. \cite{Nielsen2009}, which allows to detect and correct errors. It relies on encoding the information present in a quantum system into a larger Hilbert space. In practice, this encoding procedure requires both the ability to perform quantum operations below a certain error threshold and control of sizable quantum systems \cite{CaLi2021}, for example, the authors of  \cite{FoMa2012} showed that the surface error correcting code requires a thousand qubits per logical qubit to perform Shor's algorithm. Despite these significant experimental challenges, a recent experimental implementation of error correction was done in \cite{AcAl2023}.

The field quantum Error Mitigation (EM) provides a set of alternative strategies to reduce the effects of noise that are effective on the currently available NISQ platforms \cite{EnCa2021,CaLi2021}. These methods aim to use a limited amount of quantum operations and ancillary qubits supplemented by classical post-processing on the final measurement outcomes. 
Extrapolation methods learn the dependence of the measurement outcomes on the noise strength by increasing it and then extrapolate to 0 noise strength  \cite{TeBr2017,LiBe2017}. This method was experimentally implemented in \cite{DuMc2018,KaTe2019,GaPo2020}. 
In readout error mitigation, the measurement outcomes are corrected by applying a linear transformation that compensates for known measurement errors \cite{MaZo2020,ChFa2019}. 
Finally, the idea of quasi-probability methods relies on the assumption that the noise on every unitary operation is given by Completely Positive Trace Preserving (CPTP) maps \cite{CaLi2021}. Since the inverse of a CPTP map is CPTP if and only if the map is unitary, it cannot directly be implemented. However, if one is able to perform a complete set of CPTP operations, the inverse can be implemented with elements from this set weighted by a quasi probability distribution (i.e. a probability distribution which can take negative values) \cite{TeBr2017, EnSi2018,JiWa2021}, see also \cite{HuLi2020} and \cite{Ta2021,TaEn2022,HaMa2021} for results on the cost of this error mitigation. Quasi-probability has successfully been implemented on a super conducting quantum processor \cite{SoCu2019} and a Rydberg quantum simulator \cite{ZhaYa2020}. Recently, the authors of \cite{RoMa2023} developed an efficient algorithm to implement the inverse of any 2-level system CPTP map and implemented it on an IBM quantum processor.
Implementing the inverse of CPTP maps also has interest outside of error mitigation, see e.g \cite{HuKu2020}.

One of the main issues of the current quasi-probability-based error mitigation schemes is that one needs to be able to efficiently implement the CPTP maps that make up the desired inverse. In this paper, we propose a novel method which circumvents the challenge of implementing any CPTP maps. We take advantage of the well-known fact that any open quantum system dynamics always solves a time local evolution equation see e.g. \cite{ChKo2010} and put the system in contact with a specifically designed reservoir leading to additional terms in the master equation.
The terms added by the designed reservoir are of the Lindblad-Gorini-Kossakowski-Sudarshan form \cite{LinG1976,GoKoSu1976}, or Lindblad form in short. Therefore, by continuously monitoring the additional reservoir, quantum jump trajectories can be reconstructed for the system state \cite{WiMi1993,BrPe1995}. 
By applying a recent development in the field of quantum trajectory theory introduced by some of us \cite{DoMu2022,DoMu2023}, we show that the quantum trajectories generated by the designed reservoir can weighed by a pseudo-probability measure changing, on average, the additional terms in the system master equation. For a suitable choice of reservoir and pseudo-measure, the influence of the original noise source can be completely cancelled. 
Essentially, what we are presenting here is a simulation technique for generators of time local master equations with negative decoherence rates, which can be used for error mitigation by simulating a generator which cancels out the generator of an existing bath. Recently, the authors of \cite{GuJo2023} implemented the opposite scheme. They simulated time local master equations (with positive rate functions) using quantum error mitigation.

Alternatively, the quantum trajectories generated by the designed reservoir can be simulated using the formalism developed in \cite{LloVio2001} to simulate the evolution of time-local master equations of the Lindblad form in short. Again, relying on \cite{DoMu2022,DoMu2023}, we extend the method by \cite{LloVio2001} to general time-local master equations and in this way it can be used to replace the engineered reservoir to perform the error mitigation. The simulation of \cite{LloVio2001} scheme was experimentally implemented in \cite{SchMu2013}.


\section{Error Mitigation with Quasi Probabilities}
The error mitigation methods presented in \cite{TeBr2017, EnSi2018,SuYu2021} assume that the noise on quantum operations can be modelled by a completely positive map $E$ which we know. Counteracting $E$ essentially relies on having access to a complete set of completely-positive trace-preserving maps $S= \{ B_k\}_k$ such that 
\begin{align*}
    E^{-1} = \sum_k q_k\, B_k
\end{align*}
The normalisation of the c-numbers $q_k\in \mathbb{R}$ ensures trace  preservation $\sum_k q_k=1$. As the inverse of a completely positive map is generically only completely bounded \cite{JoKrPa2009}, the $q_k$'s can take negative values and consequently the instrument \cite{BaLu2005} associated to map on operators $E^{-1}$ specifies a pseudo probability distribution. Let us define $C = \sum_k |q_k|$, $p_k = q_k/C$ and $s_k = \textrm{sign}(q_k)$ such that we can rewrite the the above equation as
\begin{align*}
    E^{-1} = C \sum s_k \, p_k \, B_k.
\end{align*}
in terms of the probability distribution $\{p_k\}_k$ and $C$ is called the \textit{cost} of the quantum error mitigation.
The expectation value of an observable $O$ for the action of $E$ on a state $\rho$ is then computed by the formula
\begin{align}\label{eq:postprocessing}
    \textrm{tr}(O\, E^{-1}(\rho)) = C \sum s_k \, p_k \, \textrm{tr}(O\,B_k(\rho))
\end{align}
The above equation explains how the expectation value can be measured in practice. First an operator $B_k$ is drawn from the probability distribution $\{p_k\}_k$. Then $B_k$ is applied to the quantum state and the observable $O$ is measured. The final result is reweighed by $C$, multiplied by the sign $s_k$ and summed to the estimate of $\textrm{tr}(O\, E^{-1}(\rho))$.

\section{Error mitigation with quantum trajectories}\label{sec:error_mit_mart}
In this section we develop our error mitigation technique based on coupling the noisy system to an additional reservoir and reweighing the resulting trajectories with the influence martingale \cite{DoMu2022}, a pseudo-probability measure. Before going to the error mitigation, we dedicate a subsection to introduce quantum trajectories and the influence martingale.

\subsection{Simulating a non-Markovian Reservoir with Quantum Trajectories}\label{sec:martingale}
In this subsection we provide a short introduction to quantum trajectories and the influence martingale, see Appendix \ref{app:mart} for more technical details.
We consider a system in contact with a bath, such that the effective evolution  of the system state $\rho(t)$ is described by the master equation 
\begin{align}\label{eq:master}
    \frac{d}{dt}\rho(t) = -i [H(t),\rho(t)] + \mathcal{L}_t (\rho(t))
\end{align}
where $\mathcal{L}_t$ is a dissipator fully characterised by the set $\{L_k,\gamma_k(t)\}$ of Lindblad operators $L_k$ and jump rates $\gamma_k(t)\geq 0$ 
\begin{align}\label{eq:dissipator}
    \mathcal{L}_t(\rho) = \sum_k \gamma_k(t) \left(L_k\rho L_k^\dagger -\frac{1}{2}\{L_k^\dagger L_k ,\rho\}\right).
\end{align}
Note that the above master equation is of the Lindblad form if and only if the rates $\gamma_k(t)$ are positive.
When the bath is continuously monitored, quantum jump trajectories for the system state vector $\psi(t)$ can be reconstructed \cite{WiMi1993,BrPe1995}. This procedure is also called unravelling in trajectories. These trajectories are fully characterised by a set of jump times and jump operators $\{t_j,L_{k_j}\}$ which means that at the times $t_j$ the system state vector $\psi(t)$ made a jump described by the operator $L_{k_j}$:
\begin{align}
    \psi(t) \xrightarrow[\text{at } t_j]{\text{jump}} \frac{L_{k_j}\psi(t)}{\|L_{k_j}\psi(t)\|}.
\end{align}
The solution $\rho(t)$ to the master equation \eqref{eq:master} is then reconstructed by taking the average $\mathbb{E}$ of the states $\psi(t)\psi^\dagger (t)$ \cite{GaPaZo1992,DaCa1992,Carmichael1993}
\begin{align}\label{eq:solME}
    \rho(t) = \mathbb{E}(\psi(t)\psi^\dagger(t))
\end{align}
Recently \cite{DoMu2022,DoMu2023} some of us introduced a martingale method, called the influence martingale, which provides a conceptually simple and numerically efficient avenue to unravel completely bounded maps on operators.
The influence martingale pairs the evolution of a master equation with dissipator $\mathcal{L}_t$ \eqref{eq:dissipator} with positive rate functions $\gamma_k(t)$ to a dissipator $-\tilde{\mathcal{L}}_t$ characterised by $\{L_k,\Gamma_k(t)\}_k$
\begin{align}\label{eq:dissipator_NM}
    -\tilde{\mathcal{L}}_t(\rho) = -\sum_k \Gamma_k(t) \left(L_k\rho L_k^\dagger -\frac{1}{2}\{L_k^\dagger L_k ,\rho\}\right).
\end{align}
with non-sign definite decoherence rates $-\Gamma_k(t)$ if there exist a positive function $m(t)$ such that
\begin{align}\label{eq:relation}
    \gamma_k(t) + \Gamma_k(t) = m(t),\quad \forall k.
\end{align}
This pairing between dissipators is achieved with the influence martingale $\mu(t)$ which is designed to follow evolution of its corresponding quantum trajectory. For a quantum trajectory with jumps $\{t_j,L_{k_j}\}$, the influence martingale at time $t$ equals 
\begin{align}\label{eq:martingale}
    \mu(t) = \exp\left( \int_0^t ds\, m(s) \right)\prod_{j,
    \, t_{j}\leq t}\left(\frac{-\Gamma_{k_{j}}(t_{j})}{\gamma_{k_{j}}(t_{j})}\right).
\end{align}
The solution to the master equation 
\begin{align}\label{eq:master_neg}
    \frac{d}{dt}\bar{\rho}(t)= -i[H(t),\bar{\rho}(t)] - \tilde{\mathcal{L}}_t(\bar{\rho}(t))
\end{align}
where $\tilde{\mathcal{L}}_t$ has jump operators and rates $\{L_k,\Gamma_k(t)\}$, is then obtained by weighing the average over all trajectories $\mathbb{E}$ with $\mu(t)$ for each trajectory. Concretely, the solution to eq. \eqref{eq:master_neg} is given by
\begin{align}\label{eq:statebar}
    \bar{\rho}(t) = \mathbb{E}(\mu(t)\sigma(t)).
\end{align}

As outlined above, the quantum trajectories from one reservoir can be used to simulate the dynamics of another by reweighing stochastic averages with the influence martingale. In this sense, the influence martingale has the role of a quasi-probability measure in a post-processing procedure. Computing the expectation value of an observable $O$ with eq. \eqref{eq:statebar} gives $\textrm{tr}(O\rho(t)) = \mathbb{E}(\mu(t) \textrm{tr}(O\sigma(t)))$. This expression clearly resembles the error mitigation via quasi-probability methods given in eq. \eqref{eq:postprocessing}.

\subsection{Error Mitigation}\label{sec:errorMit}
We consider a system which undergoes a unitary evolution governed by a Hamiltonian $H(t)$. The free system evolution is disturbed by a bath which leads to a dissipator $\tilde{\mathcal{L}}_t$ of the form \eqref{eq:dissipator_NM} with a set of Lindblad operators and decoherence rates $\{L_k,\Gamma_k(t)\}_k$.
The system evolution is thus described by the master equation 
\begin{align}
    \frac{d}{dt}\rho(t) = -i[H(t),\rho(t)] + \tilde{\mathcal{L}}_t(\rho(t)).
\end{align}

We aim to cancel the influence of the reservoir by introducing another specifically engineered reservoir, which leads to an extra dissipator $\mathcal{L}_t$ with $\{L_k,\gamma_k(t)\}_k$ where the rates $\gamma_k(t)\geq0$. The rates are chosen such that there is a function $m(t)$ such that the relation \eqref{eq:relation} holds true.
The resulting system evolution is 
\begin{align}\label{eq:me_full}
        \frac{d}{dt} \rho(t) = -i [H(t),\rho(t)] + \tilde{\mathcal{L}}_t(\rho(t)) + \mathcal{L}_t(\rho(t))
\end{align}

\begin{figure}
    \centering
    \includegraphics{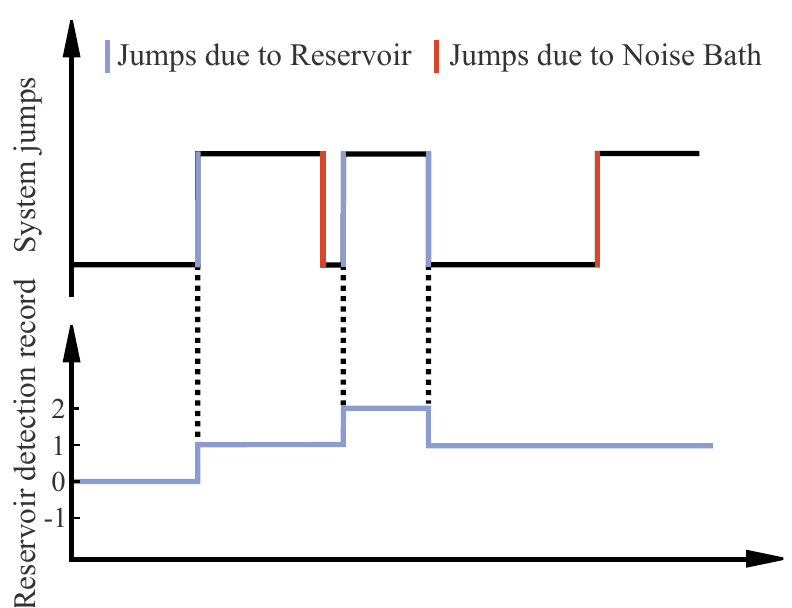}
    \caption{Illustration of a quantum jump trajectory of a quantum system (top) and the measurement record in the engineered reservoir (bottom).}
    \label{fig:traj}
\end{figure}

If the decoherence rates $\Gamma_k(t)$ of the noise bath dissipator $\tilde{\mathcal{L}}_t$ are all positive definite the full master equation \eqref{eq:me_full} can be unravelled in quantum jump trajectories. The jumps of the system are then caused by both the engineered reservoir and the noise source, in Fig. \ref{fig:traj} the to curve illustrate such a system jump trajectory. 
We assume, however, that we are only able to continuously measure the engineered reservoir. Therefore we only obtain the measurement record of the reservoir, the bottom curve in Fig. \ref{fig:traj}.
With this measurement record we obtain the set of jump times and jump operators $\{t_j,L_{k_j}\}_j$ for the system state $\psi(t)$ caused by the reservoir. With this set, we construct the martingale as in the last section \eqref{eq:martingale} and reweigh the trajectories by it.

\begin{figure}
    \centering
    \includegraphics[scale=1.3]{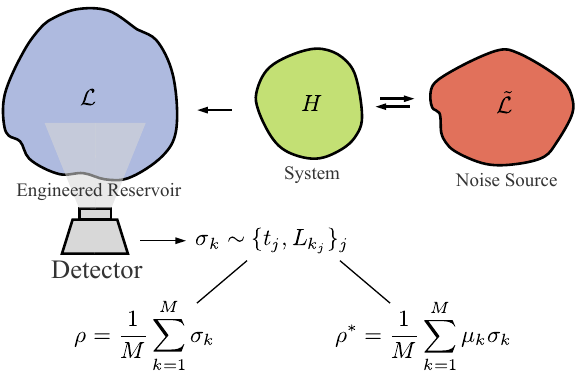}
    \caption{Our proposed simulation scheme. The system is in contact with a noise source which adds the dissipator $\tilde{\mathcal{L}}$ in its evolution. The system is brought in contact with an additional engineered reservoir which leads to the additional dissipator ${\mathcal{L}}$. The engineered reservoir is constantly observed such that trajectories for the system state $\sigma_k$ can be generated. By post selection with the influence martingale $\mu(t)$ \eqref{eq:martingale} the state $\rho^*$ can be constructed which solves the free evolution \eqref{eq:pureEvol}.}
    \label{fig:cancelling}
\end{figure}

Our error mitigation strategy is then as follows:
\begin{itemize}
    \item Bring the system in contact with an additional reservoir as illustrated in Fig. \ref{fig:cancelling} which leads to an extra dissipator ${\mathcal{L}_t}$ in the master equation with rates and jump operators $\{\gamma_k(t),L_k\}_k$. The resulting evolution is given by equation \eqref{eq:me_full}.

    \item By continuously monitoring the additional reservoir, construct a measurement record ${t_j , L_{k_j} }_{j}$ with jumps caused by Lindblad operators $L_{k_j}$ at times $t_{j}$.

    \item Weigh the trajectories by the appropriate martingale function \eqref{eq:martingale}, such that the averaged state $\rho^*(t) = E(\mu(t) \psi(t)\psi^\dagger(t))$ solves the purely unitary evolution 
    \begin{align}\label{eq:pureEvol}
        \frac{d}{dt}\rho^*(t) = -i [H(t),\rho^*(t)].
    \end{align}
\end{itemize}

The quantum error mitigation cost, as defined in \cite{SuYu2021} is given in terms of the martingale by
\begin{align*}
    C(t) &= E(|\mu(t)|) \\
    &\leq \exp\left(\int_0^t ds \left[ \min_k(-\Gamma_k(s))- |\min_k(-\Gamma_k(s))| \, \right]\right).
\end{align*}
The bound for the cost bears similarity to the expression for the cost obtained in \cite{HaMa2021}. In fact, by making the same assumption on the Lindblad operators (i.e. $L_k^\dagger L_k = \mathbb{I},\, \forall k$) as \cite{HaMa2021} recover the same expression of for cost, see Appendinx \ref{app:EM}. Note that the above bound equals 1 when all decoherence rates $\Gamma_k(t)$ are negative definite functions. In this case the $-\tilde{\mathcal{L}}_t$ is of the Lindblad form and thus the noise can be cancelled without the need implementing a quasi-probability distribution. 

In the above presentation we have made the assumption that adding the reservoir, a second bath, results in an extra dissipative term in the master equation. This is justified when the system and both bath are initially in a product states and the coupling to the baths is sufficiently weak such that the weak coupling limit can be applied. For an overview of the weak coupling limit see e.g. \cite{RiHu2012,BrPe2002} and for a discussion on the applicability of master equations in quantum computing models see \cite{McCru2020}.

\subsection{Example: Anisotropic Heisenberg Model}
This model is described by the Hamiltonian
\begin{align}\label{eq:hamHeis}
    H =& J\sum_{\langle ij\rangle}[(1+\gamma) \sigma_x^{(i)}\sigma_x^{(j)}+(1-\gamma) \sigma_y^{(i)}\sigma_y^{(j)}+  \sigma_z^{(i)}\sigma_z^{(j)}]\nonumber\\
    &-\gamma h \sum_{i=1}^4\sigma_y^{(i)},
\end{align}
where the sum over $\langle ij\rangle$ denotes a nearest interaction between the 4 spins ordered on a $2\times 2$ lattice and $\sigma^{(i)}_{x,y,z}$ are the Pauli operators acting on the $i$-th site. 

The qubits all experience local relaxation and dephasing noise, respectively described by the dissipators $\mathcal{L}_R$ and $\mathcal{L}_D$, thus $\tilde{\mathcal{L}}=\mathcal{L}_R+\mathcal{L}_D$. The dissipators are of the form \eqref{eq:dissipator} where $\mathcal{L}_R$ is characterised by the set of rates and Lindblad operators $\{\Gamma_R, \sigma_-^{(i)}\}_{i=1}^4$ and $\mathcal{L}_D$ by $\{\Gamma_D, \sigma_z^{(i)}\}_{i=1}^4$. For our model we take $\Gamma_R=\Gamma_D= 0.001 J$. 

\begin{figure}
    \centering
    \includegraphics{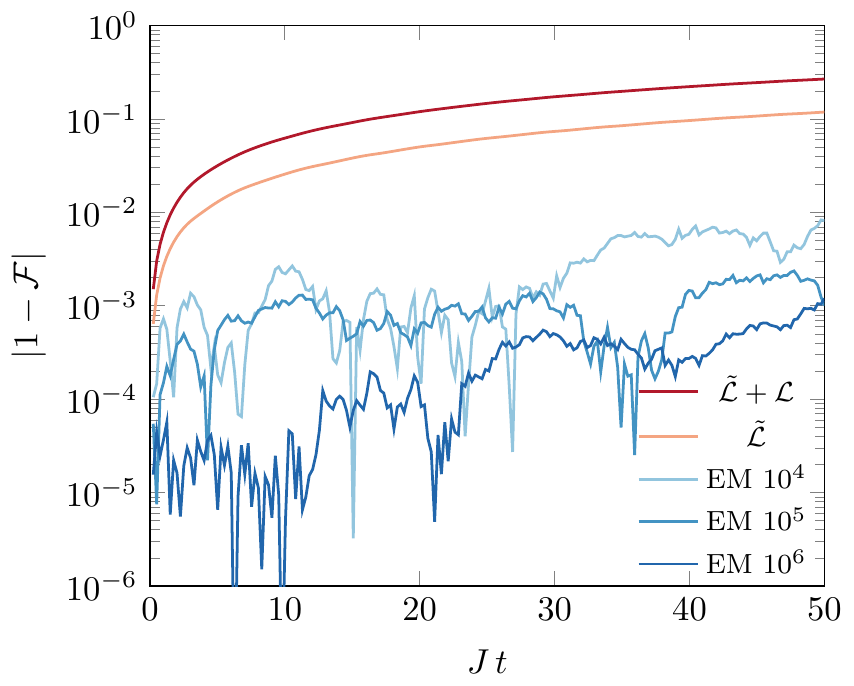}
    \caption{Fidelity $F$ of the pure unitary 4 qubit evolution governed by \eqref{eq:hamHeis} with the disturbed and error mitigated evolutions. The pure evolution is disturbed by the dissipator $\tilde{\mathcal{L}} = \mathcal{L}_R + \mathcal{L}_D$, the error mitigation leads to the extra dissipator ${\mathcal{L}}$ worsening the outcome. The blue curves show the fidelity after error mitigation for $10^4$,  $10^5$ and $10^6$ trajectories. }
    \label{fig:Fidelities}
\end{figure}

Fig. \ref{fig:Fidelities} shows illustrates the performance of our proposed error mitigation scheme by plotting $|1-\mathcal{F}|$, where $\mathcal{F}$ is the fidelity
\begin{align*}
    \mathcal{F}(\rho,\sigma)= \left(\textrm{tr} \sqrt{\sqrt{\rho}\sigma\sqrt{\rho}}\right).
\end{align*}
The (smooth) light red line shows the change in fidelity between the between target state with unitary evolution governed by $H$ \eqref{eq:hamHeis} and the disturbed evolution with the extra dissipator $\tilde{\mathcal{L}} = \mathcal{L}_R+\mathcal{L}_D$. The (smooth) dark red line shows the fidelity between the target state and the state with both the dissipator $\tilde{\mathcal{L}} $ due to the noise source and the dissipator $\mathcal{L}$ due to the additional reservoir which will be used for error mitigation.
The (noisy) blue lines show the fidelity with the martingale-based error mitigation which realises $-\bar{\mathcal{L}}$ by averaging over $10^4, \, 10^5$ and $10^6$ generated trajectories.
Let $\mathcal{F}_E$ be the fidelity of the target unitary evolution with the error correction and $\mathcal{F}_{\tilde{\mathcal{L}}}$ with the unitary evolution disturbed by $\mathcal{L}$. The difference $\log_{10}(|1-\mathcal{F}_E|)-\log_{10}(|1-\mathcal{F}_{\tilde{\mathcal{L}}}|)$ averaged over the time interval in Fig. \ref{fig:Fidelities} is ${-1.5}$, ${-1.8}$ and ${-2.6}$ for $10^4$, $10^5$ and $10^6$ trajectories, respectively.

\begin{figure}
    \centering
    \includegraphics[scale=0.92]{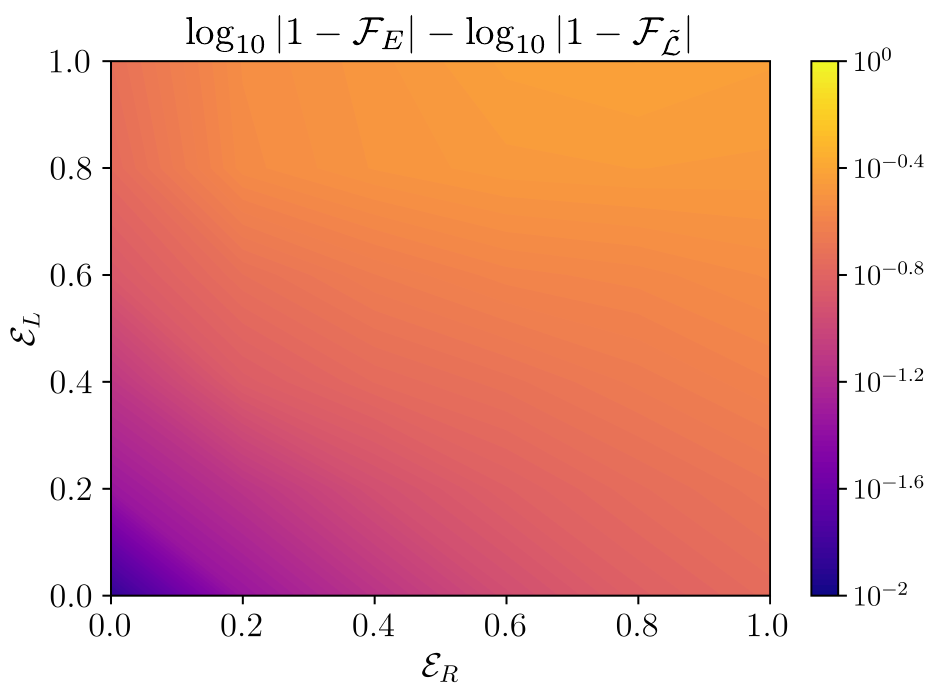}
    \caption{Improvement of the fidelity at the final time $J\,t=50$ with quantum error mitigation including errors. $\mathcal{F}_{\tilde{\mathcal{L}}}$ denotes the fidelity of the state freely evolving with the Hamiltonian $H$ \eqref{eq:hamHeis} with the state disturbed by $\tilde{\mathcal{L}} = \mathcal{L}_R+\mathcal{L}_D$ and $\mathcal{F}_E$ with the error mitigated state with errors. $\mathcal{E}_L$ gives the strength of the error on the Lindblad operators and $\mathcal{E}_R$ the strength of the errors on the rates in $\mathcal{L}$. Error mitigation is done with $10^4$ trajectories and the values of the plotted function are averaged over 10 realisations of the noise.}
    \label{fig:error_rates}
\end{figure}

We consider two types of errors in the quantum error mitigation
\begin{enumerate}
    \item[a.] \textbf{Errors in correctly identifying the Lindblad operators $L_k$ and the decoherence rates $\Gamma_k$}: We model the errors on the Lindblad operators $L_k + \eta_k \, J_k$ where $\eta_k \in [0,\mathcal{E}_L]$ and $J_k$ a matrix with uniform random entries between 0 and 1. $J_k$ only acts on the site on which $L_k$ acts, e.g. for $\sigma_x^{(i)}$ the noise operator only acts on the $i$-th site. The errors on the rates are modelled as  $\Gamma_k + \delta_k \Gamma_k$ where $\delta_k\in [0,\mathcal{E}_R]$.
    \item[b.] \textbf{Errors in correctly identifying the jump times of quantum trajectories  $\{t_j,L_{k_j}\}_j$}: The errors on the jump times are modelled by drawing a random number from $\epsilon_{j} \in[-\mathcal{E}_T/2,\mathcal{E}_T/2]$ and shifting the jump timings by $\{t_j + \epsilon_j \Gamma^{-1},L_{k_j}\}_j$.
\end{enumerate}
Fig. \ref{fig:error_rates} shows the improvement of the fidelity between the free evolution and the evolution disturbed by the noise source $\mathcal{F}_{\tilde{\mathcal{L}}}$ and the fidelity with error mitigation with errors $\mathcal{F}_{E}$ as described in point a. above, averaged over 10 realisations of the errors. Even for errors of the order of the norm of the Lindblad operators and the jump rates, the error mitigation still gives improvement in the fidelity. In Fig. \ref{fig:error_rates} we show the fidelity for the errors defined in point b. as a function of the size of the errors, again averaged over 10 realisations of the errors.

\begin{figure}
    \centering
    \includegraphics{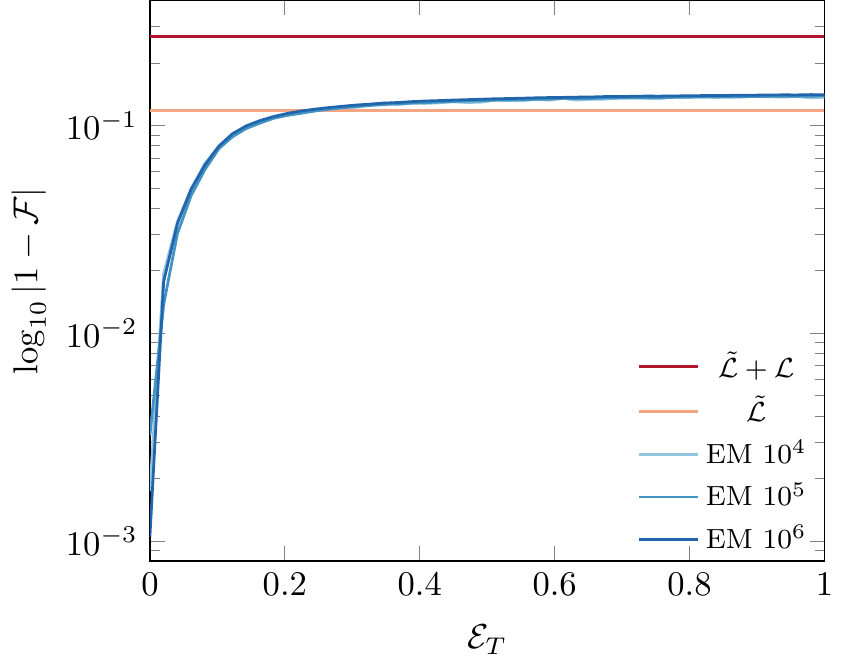}
    \caption{Final state fidelity as a function of the error on jump detection times. The three (overlapping) full lines show the performance of the error mitigation for the state at time $Jt=50$ for $10^4$, $10^5$ and $10^6$ trajectories. The result is the average over 10 realisations of the noise. The (full) light red line shows the value of the final state fidelity under the noise described by the dissipator $\tilde{\mathcal{L}}$.}
    \label{fig:error_times}
\end{figure}

\section{Error mitigation with Simulated Quantum Trajectories}\label{sec:LV}
In the last section, we showed that performing post-processing on the quantum trajectories of a specially engineered reservoir can be used as a quantum error mitigation technique to cancel out the influence of a noise source. Constructing such a reservoir in full generality can be challenging. Luckily, we do not require an actual reservoir but just that the system undergoes the appropriate quantum jump trajectories.

In this section, we show that the these trajectories can be simulated using the scheme developed by Lloyd and Viola in \cite{LloVio2001} by letting the system interact with one ancillary qubit and measuring that qubit. Using this scheme for error mitigation relies on the assumption that interaction with the ancilla and measurement happen on a much faster time scale than the interaction with the reservoir. The error mitigation scheme shown in Fig. \ref{fig:Llyod-Viola} then consists splitting the total interaction time with the bath up into intervals of size $\Delta t$. Before each of these interval, an interaction with the ancilla ($U_{LV}$) takes place, which negates the environmental influence of the coming step.

\subsection{Simulating Quantum Trajectories for Lindblad master equations}\label{sec:LValg}
In a physical system, Lindblad dissipators $\mathcal{L}_t$ \eqref{eq:dissipator} can be realised with the simulation scheme for Lindblad dissipators proposed in \cite{LloVio2001}. The simulation scheme essentially realises the stochastic quantum jump trajectories which we discussed in Section \ref{sec:martingale}. These trajectories are generated by repeatedly letting the system interact with an ancilla qubit and measuring that ancilla. The Lindblad evolution generated by ${\mathcal{L}}_t$ is then obtained by averaging over the generated trajectories. 
Therefore, the influence martingale \eqref{eq:martingale} can be used to weigh the trajectories just as in the previous section, such that that time local master equations with non-positive-definite decoherence rates.

We present here a simplified version of the scheme of \cite{LloVio2001} in the case of one Lindblad operator $L$ with $L^\dagger L = \mathbb{I}$ and the polar decomposition $L= U\, X$ where $U$ is a unitary and $X$ a positive operator. To simulate a generator with Lindblad operator $L$ and jump rate $\gamma$ for time steps $\Delta t = t_C/\alpha$ the system is coupled with an ancillary qubit through the Hamiltonian
\begin{align}\label{eq:hamLV}
    H_{LV} = \sqrt{\frac{\gamma}{\alpha t_C}} X \otimes \sigma_x
\end{align}
for a time $t_C$. Let the initial condition be $\rho(0) = \rho_S\otimes |0\rangle\langle0|$, where $\rho_S$ is the system initial state and $\sigma_z |0\rangle= -|0\rangle$. A straightforward computation gives an explicit expression for the unitary time-evolution operator of the composite system
\begin{align}\label{eq:unitaryLV}
    e^{-iH_{LV}t_C} =& \cos\left(\sqrt{\frac{\gamma}{\alpha t_C}} t_C X\right)\otimes\mathbb{I}- i \sin\left(\sqrt{\frac{\gamma}{\alpha t_C}}t_C X\right)\otimes \sigma_x.
\end{align}
Let $\rho(t_C)$ be the composite state after an interaction time $t_C$
\begin{align}
    \rho(t_C) = e^{-i H_{LV} t_C} \rho(0)e^{i H_{LV} t_C}.
\end{align}
After this interaction, the ancillary qubit is measured in the $|0\rangle,\, |1\rangle$ basis, with
\begin{align}
    \sigma_z|0\rangle = -|0\rangle\quad \textrm{and}\quad\sigma_z|1\rangle = |1\rangle\nonumber.
\end{align} We assume that $\sqrt{\frac{\gamma}{\alpha t_C}} t_C$ is small enough such that we can expand all expressions below up to second order in $\sqrt{\frac{\gamma}{\alpha t_C}} t_C$. The state $|0\rangle$ is measured with probability 
\begin{align}
    p_0 = 1-\gamma \Delta t\,\textrm{tr}(L^\dagger L \rho_S)
\end{align}
and the system is in the state
\begin{align}\label{eq:meas0}
\frac{\textrm{tr}_2(\rho(t_C)\,\mathbb{I}\otimes|0\rangle\langle0|) }{p_0}=  \frac{\rho_S - \frac{\gamma \Delta t}{2} \{L^\dagger L,\rho_S\}}{1-\gamma \Delta t\,\textrm{tr}(L^\dagger L \rho_S)},
\end{align}
where $\textrm{tr}_2$ is the partial trace over the qubit Hilbert space and we used that $L=U\,X$, with $U$ a unitary operator and thus $X^2  = L^\dagger L$.
After measuring 1 with probability 
\begin{align*}
    p_1= \gamma \Delta t\,\textrm{tr}(X^\dagger X \rho_S) 
\end{align*}
the unitary $U$ is multiplied to the system such that it is in the state
\begin{align}\label{eq:meas1}
    U\frac{\textrm{tr}_2(\rho(t_C)\,\mathbb{I}\otimes|1\rangle\langle1|) }{p_1}U^\dagger=  \frac{L\rho_SL^\dagger}{\textrm{tr}(L^\dagger L \rho_S)}
\end{align}

After averaging over both measurement outcomes the system is in the state
\begin{align}
    \rho_S(\Delta t) = p_0  \frac{\textrm{tr}(\rho(t_C)\,\mathbb{I}\otimes|0\rangle\langle0|) }{p_0}+p_1 \frac{\textrm{tr}(\rho(t_C)\,\mathbb{I}\otimes|1\rangle\langle1|) }{p_1}
\end{align}
and we find that
\begin{align}
    \frac{\rho_S(\Delta t)-\rho_S}{\Delta t} = \gamma\left(L\rho_SL^\dagger-\frac{1}{2}\{L^\dagger L,\rho_S\}\right)
\end{align}
and have thus simulated one time step $\Delta t$ of the evolution of a master equation with Lindblad operator $L$ and rate $\gamma$.

Let us take a closer look at the states after measurement outcomes 0 \eqref{eq:meas0} and 1 \eqref{eq:meas1}. For small time steps $\Delta t$, the probability to measure 0 is of order $O(1 - \Delta t)$ and the system state undergoes a small change of order $O(\Delta t)$. For small $\Delta t$, mostly outcome 0 is measured and the state evolves continuously \eqref{eq:meas0} on the rare events that 1 is measured, the system undergoes a sudden change, a quantum jump. A quantum trajectory is thus generated by $N$ consecutive interactions with the ancillary qubit and is fully characterised by set of jump times $\{t_j=n_j\Delta t,L\}$ where $n_j\in \mathbb{N}, \, \forall j$. By averaging over all trajectories, a system evolution of a master equation with Lindblad operator $L$ and rate $\gamma$ is simulated for a time $t=N\Delta t$. In the limit of $\Delta t \downarrow 0$ these trajectories are exactly the quantum jump trajectories discussed in Section \ref{sec:martingale} of a dissipator with Lindblad operator $L$ and rate $\gamma$.

The generalisation to $K$ Lindblad operators and rates $\{L_k, \gamma_k\}_{k=1}^K$ is given in the supplementary material. This can either be done by performing $K$ consecutive qubit interactions, each time with a different operator $X_k$ and coupling strengths in $H_{LV}$ \eqref{eq:hamLV}, and measurements. On the other hand, we can resort to a probabilistic approach and for each time step draw a random Lindblad operator $L_k$ with probability $1/K$ and choose $X_k$ and the required coupling strength in the Hamiltonian \eqref{eq:hamLV} to implement one step with $L_k$ and $\gamma_k$. As in the single Lindblad operator case, the trajectory generated by $N$ interactions with the ancilla is characterised by the set of jump times and Lindblad operators $\{t_j=n_j\Delta t,L_{k_j}\}$ and in the limit of $\Delta t\downarrow 0$ they recover the quantum jump trajectories unravelling a dissipator with $\{L_k, \gamma_k\}_{k=1}^N$.

A Hamiltonian term $H(t)$ in the master equation can be simulated as well. This can be done by making use of Trotter's formula. After the $k$-th time step of size $\Delta t$ is simulated, the unitary operator $\exp(-iH(k\Delta t) \Delta t)$ is applied to the state of the system. In the limit of $\Delta t\downarrow 0$ the system state averaged over all trajectories solves the master equation with the simulated dissipator and the Hamiltonian $H(t)$.

\subsection{Simulating Quantum Trajectories for general master equations}\label{sec:LValg}
Let us now perform additional post-processing after the measurement on the ancilla qubit to implement the martingale pseudo-measure \eqref{eq:martingale}. Depending on the measurement outcome, we multiply $\mu$
\begin{align}
    \mu = \begin{cases}
     1 + m {\Delta t} = 1 + m {\Delta t}\, L^\dagger L\,\, & \textrm{0 measurement}\\
     +\frac{\gamma-m}{m}\,\, & \textrm{1 measurement}
    \end{cases}
\end{align}
After averaging over the measurement results we find 
\begin{align}
    \bar{\rho}_S(\Delta t) =& p_0 ( 1 + m {\Delta t}\, L^\dagger L)  \frac{\textrm{tr}(\rho(t_C)\,\mathbb{I}\otimes|0\rangle\langle0|) }{p_0}\nonumber\\&-p_1\frac{\gamma-m}{\gamma} \frac{\textrm{tr}(\rho(t_C)\,\mathbb{I}\otimes|1\rangle\langle1|) }{p_1}
\end{align}
up second order in  $\sqrt{\frac{\gamma}{\alpha t_C}} t_C$, the above equation leads to
\begin{align}
    \frac{\bar{\rho}({\Delta} t)-\rho_S}{{\Delta} t} =- \Gamma \left(L\rho_SL^\dagger-\frac{1}{2}\{L^\dagger L,\rho_S\}\right)
\end{align}
which simulates one time step of a master equation with a dissipator with Lindblad operator $L$ and decoherence rate $-\Gamma = \gamma-m$ which is not necessarily positive definite.

After $N$ interactions with the ancilla, a quantum trajectory $\{t_j=n_j\Delta t,L\}$ is generated. The total multiplicative factor equals
\begin{align}
    \mu(N t) = \prod_{n,\, n\leq N}\left(1 + m\Delta t\right) \prod_{j,\, n_j\leq N} \left(\frac{-\Gamma}{\gamma}\right),
\end{align}
which in the limit of $\Delta t\downarrow 0$ recovers the influence martingale pseudo measure \eqref{eq:martingale} in the case of one Lindblad operator.

For $K$ Lindblad operators with rates $\gamma_k$ and decoherence rates $\Gamma_k$, the global multiplicative factor equals
\begin{align}\label{eq:martingale_discrete}
    \mu(t) = \prod_{n,\, n\Delta t\leq t}\left(1+ m\Delta t\right) \prod_{j,\, n_j\Delta t\leq t} \left(\frac{-\Gamma_{k_j}}{\gamma_{k_j}}\right),
\end{align}
which, again, recovers the martingale \eqref{eq:martingale} in the limit $\Delta t\downarrow 0$.
\begin{figure}
    \centering
    \includegraphics{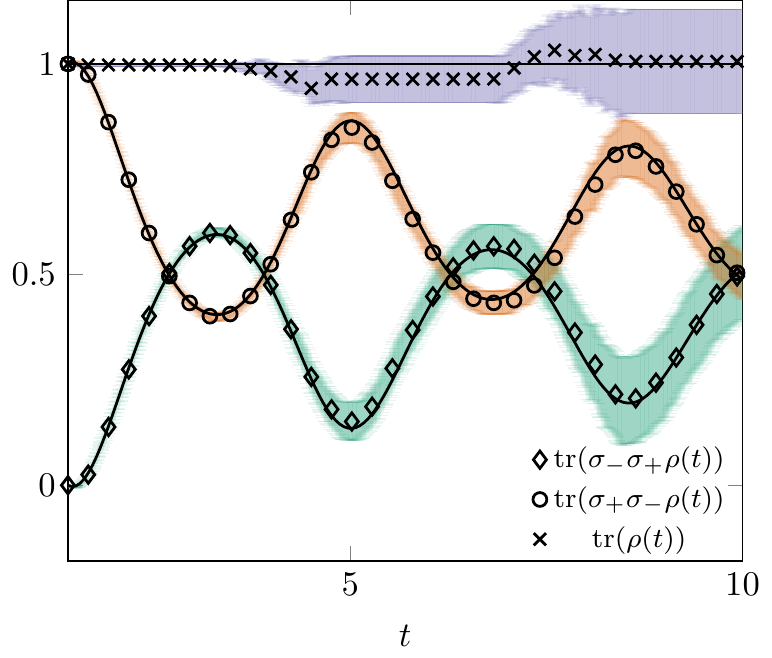}
    \caption{Computation of the solution of the 2-level system master equation \eqref{eq:masterJQ} using the simulation of quantum trajectories developed in Sec. \ref{sec:LValg}. The initial 2 level system state equals $\rho(t_i) = (\mathbb{I}+\sigma_z)/2$. The trace of the state is shown by the crosses and the diamonds and circles show expectations normalised by the trace. The full lines show the numerical integration of \eqref{eq:masterJQ}. }
    \label{fig:LV_JQ}
\end{figure}
Fig. \ref{fig:LV_JQ} shows the simulation of the time evolution of a 2-level atom in a photonic bandgap \cite{JoQu1994}. The evolution of the 2-level is governed by the master equation 
\begin{align}\label{eq:masterJQ}
    \frac{d}{dt}\rho(t) =& i\frac{S(t)}{2}[\sigma_+\sigma_-,
    \rho(t)]\nonumber\\& + \Gamma(t)\left(\sigma_-\rho(t)\sigma_+ - \frac{1}{2}\{\sigma_+\sigma_-,\rho(t)\} \right)
\end{align}
where
\begin{align}
    S(t) = -2\,\textrm{Im}\frac{\frac{d}{dt}c(t)}{c(t)},\quad \Gamma(t) = -2\,\textrm{Re}\frac{\frac{d}{dt}c(t)}{c(t)}
\end{align}
and $c(t)$ is given by eq. (2.21) in \cite{JoQu1994} with $\beta=-\delta$.

\subsection{Error mitigation}

\begin{figure}
    \centering
    \includegraphics[scale=1.4]{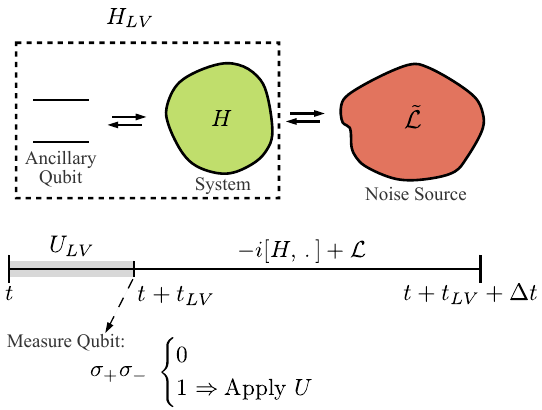}
    \caption{Error mitigation scheme with simulated trajectories. The evolution is divided into steps of size $\Delta t$. Before each step, the error mitigation is applied to cancel out the ensuing effect of the noise bath.}
    \label{fig:Llyod-Viola}
\end{figure}

We illustrate how the simulation of quantum trajectories using an ancilla qubit can be used to implement the error scheme developed in Sec. \ref{sec:errorMit}.
Figure \ref{fig:Llyod-Viola} shows how we use the extension of \cite{LloVio2001} to cancel the dissipator $\bar{\mathcal{L}}$ with Lindblad operates and jump rates $\{L_k,\Gamma_k(t)\}_k$ of the unwanted noise source.

We assume that the time $t_{LV}$ to implement one time step of the trajectory simulation scheme $t_{LV}\ll |\bar{\mathcal{L}}|^{-1}$ such that the step can be implemented without disturbance of the environment. The time $t_{LV}$ consist of the time to implement the unitary evolution
of system and ancilla \eqref{eq:unitaryLV}, to perform the measurement and to implement the eventual unitary on the system depending on the measurement outcome.

We then divide the system evolution up into time intervals of length $\Delta t$. At the beginning of each time interval, we implement one step of the quantum trajectory simulation scheme introduced in the last section. 
Then, for a simulated quantum trajectory with jumps $\{t_j = n_j \Delta t, L_{k_j}\}_j$, we construct the measure \eqref{eq:martingale_discrete}.
When averaging the trajectories weighted by the martingale function defined above the evolution with dissipator $-\bar{\mathcal{L}}$ is simulated. Thus, in the limit of $\Delta t\downarrow 0$ the original dissipator due to the noise $\bar{\mathcal{L}}$ is cancelled.

\begin{figure}
    \centering
    \includegraphics{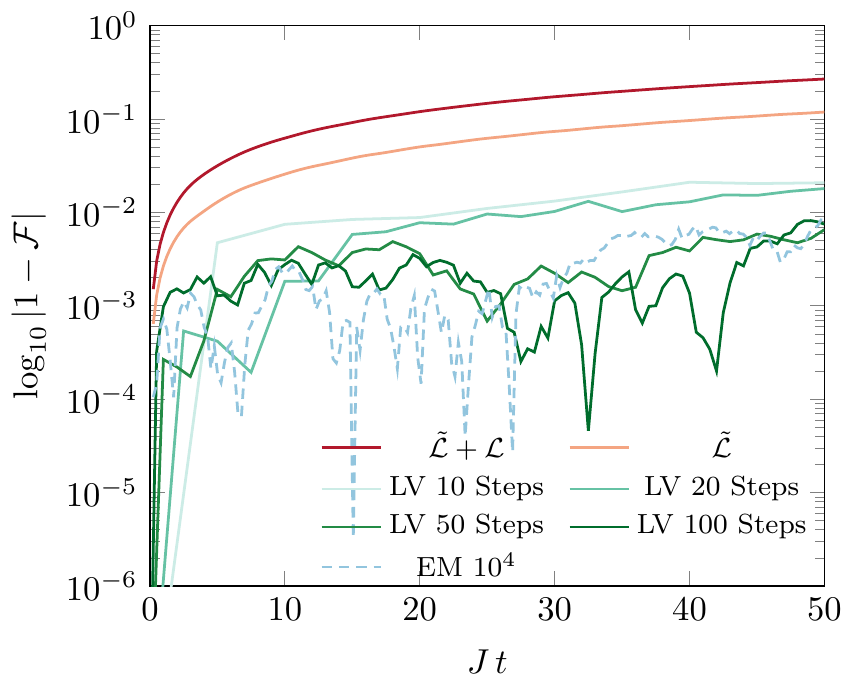}
    \caption{Fidelity $F$ of the pure unitary 4 qubit evolution governed by \eqref{eq:hamHeis} with the disturbed and error mitigated evolutions. $10^4$ trajectories were generated and total time interval is divided in different amount of a time steps, i.e. different $\Delta t$ are chosen.}
    \label{fig:FidelitiesLV}
\end{figure}

In Fig. \ref{fig:FidelitiesLV} we illustrate quantum-trajectory-simulation-based error mitigation applied to the Heisenberg model \eqref{eq:hamHeis} studied in Sec. \ref{sec:errorMit}. We see that for 50 and 100 time steps (i.e. for $\Delta t = (50/J)/$steps) the fidelity of the state evolving without noise with the error mitigated state is similar to the fidelity with the error mitigated state using an engineered reservoir studied in Sec. \ref{sec:errorMit}.

\section{Conclusion}
 We developed and demonstrated a new quantum error mitigation scheme which establishes a connection between recent results from quantum trajectory theory \cite{DoMu2022,DoMu2023} and quasi-probability based error-mitigation techniques \cite{TeBr2017, EnSi2018,JiWa2021}. Concretely, we consider a situation where the unitary evolution of a system is disturbed by a noise source whose effect can be modelled adding a dissipator to the Liouville-von Neumann equation which described the dynamics of the system. Our proposed scheme relies to bring an additional, specifically engineered, reservoir in contact with the system. By monitoring this reservoir, records of quantum jumps trajectories of the system are obtained. We showed that these quantum jump trajectories can be weighed the influence martingale, a pseudo-probability measure developed in \cite{DoMu2022,DoMu2023}, such that on average the dissipator due to the noise bath is cancelled.
 
 The quantum jump trajectories obtained by monitoring the engineered reservoir are the essential ingredient to perform our error mitigation technique. We extended the Lindblad master equation simulation technique developed by Lloyd and Viola \cite{LloVio2001} to general time local master equations. In this way, we can generate the quantum jump trajectories by bringing the system in contact with just a single ancillary qubit that is repeatedly measured.

 We illustrated our proposed error mitigation technique both based both on the engineered reservoir and the interaction with an ancillary qubit for the anisotropic Heisenberg model. We found significant improvement in fidelity with the free, unitary system evolution.

 Another application for influence martingale to error mitigation could be to use the methods presented here to implement a form of extrapolation mitigation \cite{TeBr2017,LiBe2017} to change the effects of the noise and extrapolate it to 0 strength.

\appendix 
\onecolumngrid
\section{Unravelling Time-Local Master Equations}\label{app:mart}
\subsection{Unravelling an equation of the Lindblad form}
Lindblad dynamics for a state operator $\rho(t)$ are generated by the differential equation
\begin{align}\label{eq:lindblad}
    \frac{d}{dt}\rho(t) =& - i[H(t),\rho(t)] 
    \nonumber \\& + \sum_k \gamma_k(t)\left(L_k\rho(t)L_k^\dagger -\frac{1}{2}\{L_k^\dagger L_k,\rho(t)\}\right)
\end{align}
where the jump rates $\gamma_k(t)\geq 0$. The dynamics of the Lindblad equation can also be obtained by the so-called unravelling in quantum trajectories. These quantum trajectories are generated by stochastic differential equations containing either Wiener noise or random jumps governed by Poison processes. Here we are concerned with the latter.
Consider the state vector $\psi(t)$ which solves the stochastic Schr\"{o}dinger equation 
\begin{align}\label{eq:sse}
    d\psi(t) =& -iH(t) \psi(t)\, dt
    \nonumber \\& - \frac{1}{2}\sum_k \gamma_k(t) \left( L^\dagger_k L_k  -\|L_k\psi(t)\|^2 \right) \psi(t)\, dt
    \nonumber \\& + \sum_k\left(\frac{ L_k\psi(t)}{\|L_k \psi(t)\|} -\psi(t) \right)dN_k
\end{align}
where $d\psi(t)= \psi(t+dt)-\psi(t)$ and the $dN_k$ are increments of counting processes $N_k$. The increments equal 0 when no jumps happen and 1 when a jump happens. The rates of the counting processes, conditioned on the system state are
\begin{align}
    \mathbb{E}(dN_k|\psi(t)) = \gamma_k(t) \|L_k\psi(t)\|^2 dt.
\end{align}

The solution $\rho(t)$ of the Lindblad master equation \eqref{eq:lindblad} is then obtained by the average
\begin{align}
    \rho(t) = \mathbb{E}(\psi(t)\psi^\dagger(t)).
\end{align}

Equivalent to the stochastic Schr\"{o}dinger equation for the state vector, Lindblad dynamics can be unravelled by a stochastic master equation for a state operator $\sigma(t)$ 
\begin{align}\label{eq:sme}
    d\sigma(t) =& -i [H(t),\sigma(t)] 
     \nonumber \\& -\frac{1}{2}\sum_k \gamma_k(t)\left( \{L_k^\dagger L_k,\sigma(t)\} - 2 \tr(L_k^\dagger L_k \sigma(t)) \sigma(t)\right)
     \nonumber \\& + \sum_k\left(\frac{L_k\sigma(t)L_k^\dagger}{\tr(L_k^\dagger L_k \sigma(t))} -\sigma(t)\right) dN_k.
\end{align}
It is indeed straightforward to check that setting $\sigma(t) = \psi(t)\psi^\dagger(t)$ solves the above master equation.

\subsection{Unravelling time-local master equations with the influence martingale}
Let us now introduce the martingale stochastic process $\mu(t)$ whose evolution is enslaved to the stochastic state vector evolution \eqref{eq:sse}
\begin{align}\label{eq:app_martingale}
    \begin{cases}
    d\mu(t) &= m(t) \mu(t) dt + \mu(t) \sum_k \left(\frac{\gamma_k(t)-m(t)}{\gamma_k(t)}-1\right)dN_k(t)\\
    \mu(0)&=1.
    \end{cases}
\end{align}
We then define state
\begin{align*}
    \rho'(t) = \mathbb{E}(\mu(t) \psi(t)\psi^\dagger(t))
\end{align*}
which, using the rules of stochastic calculus \cite{Jacobs2009}, can be proven to solve the master equation \cite{DoMu2022, DoMu2023}
\begin{align}\label{eq:lindblad}
    \frac{d}{dt}\rho'(t) =& - i[H(t),\rho'(t)] 
    \nonumber \\& - \sum_k \Gamma_k(t)\left(L_k\rho'(t)L_k^\dagger -\frac{1}{2}\{L_k^\dagger L_k,\rho'(t)\}\right).
\end{align}
where 
\begin{align}
    \gamma_k(t) + \Gamma_k(t) = m(t)
\end{align}
and note that since $m(t)$ is an arbitrary scalar function and therefore the decoherence rates $\Gamma_k(t))$ are not necessarily positive definite.

\section{Quantum Error Mitigation}\label{app:EM}
\subsection{Error Mitigation Scheme}
We consider a system undergoing a unitary evolution governed by a Hamiltonian $H(t)$. The system's unitary evolution is disturbed by a noise source which leads to a dissipator $\mathcal{L}$ in the system evolution
\begin{align}
    \frac{d}{dt}\rho(t) = -i[H(t),\rho(t)] + \tilde{\mathcal{L}}_t(\rho(t))
\end{align}
with 
\begin{align}
    \tilde{\mathcal{L}}_t(\rho) = \sum_k \Gamma_k(t) \left(L_k\rho L_k^\dagger -\frac{1}{2}\{L_k^\dagger L_k ,\rho\}\right).
\end{align}
We will cancel out the influence of the noise source by bringing the system in contact with a specifically engineered reservoir which leads to an additional dissipator ${\mathcal{L}}_t$ in the evolution of the system state operator 
\begin{align}\label{eq:two_baths}
    \frac{d}{dt}\rho(t) = -i[H(t),\rho(t)] + \tilde{\mathcal{L}}_t(\rho(t)) ++{\mathcal{L}}_t(\rho(t))
\end{align}
with 
\begin{align}
    \mathcal{L}_t(\rho) = \sum_k \gamma_k(t) \left(L_k\rho L_k^\dagger -\frac{1}{2}\{L_k^\dagger L_k ,\rho\}\right).
\end{align}
and positive jump rates
\begin{align}\label{eq:rates_rel}
    \gamma_k(t) + \Gamma_k(t) =m(t).
\end{align}
If the additional reservoir is continuously observed, the system evolution becomes a stochastic master equation
\begin{align}
    \frac{d}{dt}\sigma(t) =& -i[H(t),\sigma(t)] + \tilde{\mathcal{L}}_t(\sigma(t))\nonumber \\
    &-\frac{1}{2}\sum_k \gamma_k(t)\left( \{L_k^\dagger L_k,\sigma(t)\} - \tr(L_k^\dagger L_k \sigma(t)) \sigma(t)\right)\nonumber\\& + \sum_k\left(L_k\sigma(t)L_k^\dagger -\sigma(t)\right) dN_k
\end{align}
note that $\mathbb{E}(\sigma(t))$ solves \eqref{eq:two_baths}.
Let $\mu(t)$ be the influence martingale evolving according to \eqref{eq:martingale} and define the state 
\begin{align}
    \rho^*(t) = \mathbb{E}(\mu(t) \sigma(t)).
\end{align}
Again, using the rules of stochastic calculus and using the relation \eqref{eq:rates_rel}, it is possible to show that
\begin{align}
    \frac{d}{dt} \rho^*(t) &= -i[H(t),\rho(t)] + \tilde{\mathcal{L}}(\rho(t)) -\tilde{\mathcal{L}}(\rho(t))
    \\& = -i[H(t),\rho(t)]
\end{align}

\subsection{Bound on the Cost}
The cost for the quantum error mitigation which we propose relying on the influence martingale $\mu(t)$ is given by
\begin{align}
    C(t) = E(|\mu_t|)
\end{align}
using the result in \cite{DoMu2023}, and the Cauchy-Schwarz inequality $E(|\mu(t)|) \leq E(\mu(t)^2)$ (where we used that $E(1)=1)$), we find that for $m(t)= 2 \min_k(\Gamma_k(t),0) $
\begin{align}
    E(|\mu_t|)\leq \exp\left( \int_0^t ds [1 -\textrm{sign}(\min_k(\Gamma_k(t),0))] \,|\min_k(\Gamma_k(t),0)|\right).
\end{align}
Alternatively, assuming that $L_k^\dagger L_k =\mathbb{I},\, \forall k$, similar to \cite{DoMu2023}, we recover the expression for the cost of \cite{HaMa2021}
\begin{align}
    E(|\mu_t|)\leq \exp\left(\sum_k \int_0^t ds [1 -\textrm{sign}(\Gamma_k(s))] \,|\Gamma_k(s)|\right).
\end{align}

\section{Extension of Lloyd Viola}

\subsection{$N$ positive channels}\label{subsec:nPos}
Let us now consider a master equation with $N$ Lindblad operators, i.e. with a dissipator of the form \eqref{eq:dissipator} with operators and rates $\{L_k,\gamma_k\}_{k=1}^N$

For $N$ positive channels, the scheme developed in \cite{LloVio2001} requires at most $N$ consecutive measurements. Let the Lindblad operators have the polar decomposition  $L_k = U_k A_k$, where $U_k$ is a unitary and $A_k= |L_k| =\sqrt{L_k^\dagger L_k}$ a positive operator. 
Before outlining the measurement scheme, it is convenient to introduce two results.

\begin{prop}\label{prop:coupleK}
A couple of positive Kraus operators $M_{1,2}$ that satisfies the completeness relation $M_1^\dagger M_1+ M_2^\dagger M_2=\mathbb{I}$ can always be written in terms of an operator $X$
\begin{equation}\label{eq:sincos}
M_1 = \cos(X),\qquad M_2 = \sin(X).
\end{equation}
\begin{proof}
From the completeness relation 
\begin{equation}
M_1^\dagger M_1+ M_2^\dagger M_2=\mathbb{I}
\end{equation}
we find that $M_{1,2}$ are simultaneously diagonalisable. Indeed, let $U$ be the unitary that diagonalises $M_1$, written as $D_1$. Acting on both sides of the completeness relation with $U$ on the left and $U^\dagger$ on the right gives 
\begin{equation}
D_1^\dagger D_1+ U M_2^\dagger M_2U^\dagger=\mathbb{I}.
\end{equation}
Such that $U M_2^\dagger M_2U^\dagger$ is diagonal and therefore $D_2=\sqrt{U M_2^\dagger M_2U^\dagger}$ is also diagonal. Now it should be clear that $M_{1,2}$ can indeed be written as \eqref{eq:sincos}.
\end{proof}
\end{prop}

\begin{lem}\label{lem:kernel}
Let $M_{1,2}$ be positive operators. If $\psi$ is in the kernel of $\sqrt{M_1^2+M_2^2}$ then it is also in the kernel of $M_1$ and $M_2$.
\end{lem}

Let us now define the operators $B_j$ for $j=0$ to $n-1$, where $B_0=1- \frac{t_C^2}{2} \sum_k \varepsilon_k^2 A_k^\dagger A_k$ and
\begin{align}
B_j &= t_C \sqrt{\sum_{k=j}^{n} \varepsilon_k^2 A_k^\dagger A_k}\quad \text{for} \quad j\geq 1.\\
\varepsilon_k &= \frac{\sqrt{\gamma_k \Delta t}}{t_C}
\end{align}
it should be clear from their definition that the $B_j$ are positive operators. Note that this does not impose any constraints on the Lindblad rates $\gamma_k$, the interaction time $t_C$ or the simulated time-step length $\Delta t$.\\
We then implement up to $N$ consecutive quantum measurements, as outlined in Section \ref{sec:LValg}:
\begin{itemize}
\item Measurement 0: Is performed with operators $B_0$, $B_1$ which satisfy the completeness relation $B_0^\dagger B_0 + B_1^\dagger B_1 =\mathbb{I}$ (neglecting terms of order $O(\Delta t^4)$). By Proposition \ref{prop:coupleK} we know that we can implement this measurement just as outlined above. If the measurement outcome is $0$ we are done, if the outcome is $1$ we implement another measurement.
\item Measurement $j>0$: We perform a measurement with $G_0=\varepsilon_j t_C A_{j}B_j^{-1}$ and $G_1=B_{j+1}B_j^{-1}$. Note that by Lemma \ref{lem:kernel} taking the inverse makes sense since we have measured $B_j$ in the last measurement and therefore the current state of the system cannot be orthogonal to $B_j$. Furthermore Lemma \ref{lem:kernel} states that if $\psi$ is in the kernel of $B_j$ then it is also in the kernel of $B_{j+1}$ and $A_{j}$. 

The measurement operators satisfy the completeness relation
\begin{equation}
G_0^\dagger G_0 + G_1^\dagger G_1 = \mathbb{J}_j
\end{equation}
where $\mathbb{J}_j \psi=0$ when $\psi$ is in the kernel of $B_j$ and $\mathbb{J}_j \psi=\psi$ otherwise (so Proposition \ref{prop:coupleK} can still be used). By using the polar decomposition for $G_{0,1}=W_{0,1}\, |G_{0,1}|$ we can perform the measurement performed as outlined before.

If the measurement outcome is $0$ we apply the unitary $U_j$ to the system, otherwise we perform measurement $j+1$
\item Measurement $n-1$: The last measurement is performed with the operators $G_0= \varepsilon_{n-1} t_C A_{n-1}B_{n-2}^{-1}$ and $G_1=\varepsilon_{n}t_C A_{n}B_{n-2}^{-1}$. This can again be performed using the polar decomposition. For measurement outcome $0$ we perform the unitary $U_{n-1}$ and otherwise $U_{n}$.
\end{itemize}

To find the operator $X$ and corresponding coupling rate $\delta$ in the coupling Hamiltonian between system and ancilla 
\begin{align}
    H_{LV} = \delta t_C X \otimes \sigma_x,
\end{align}
let $V$ the unitary that diagonalises $|G_0|$, $D = V |G_0| V^\dagger$, $D$ diagonal. Then we impose
\begin{equation}
   \cos(\delta t_C V^\dagger X V)  \overset{!}{=} D\,.
\end{equation}
Because the right-hand side is diagonal, we can apply the inverse cosine to every single element to find the unit-trace operator $V^\dagger X V$ and along with it the factor $\delta t_C = \tr(\arccos(D))$. Applying again the unitary $V$ we obtain the operator $X$.\\
Overall the presented procedure implements the target evolution
\begin{align}
    \rho_S \,\rightarrow \, B_0 \rho_s B_0 + \sum_{k=1}^n \varepsilon_k^2 t_C^2 U_k A_k \rho_S A_k U_k^\dagger &= \rho_S + \sum_{k=1}^n  \gamma_k \Delta t \left( L_k \rho_S L_k^\dagger - \frac{1}{2} \{ L_k^\dagger L_k, \rho_S \} \right)\\
    \Rightarrow \quad \frac{\rho_S (\Delta t) - \rho_S}{\Delta t} &= \sum_{k=1}^n  \gamma_k \left( L_k \rho_S L_k^\dagger - \frac{1}{2} \{ L_k^\dagger L_k, \rho_S \} \right) 
\end{align}

\subsection{Probabilistic Approach}

Alternatively to the procedure outlined in the previous section, a master equation comprising $N$ Lindblad operators can be implemented using the scheme developed in \cite{LloVio2001} probabilistically.\\
To this end, we first of all recall the implementation of a single Lindblad operator $L=UX$, U unitary and X positive, as it is presented in Section \ref{sec:LValg}. We couple the system to an ancilla and evolve the system according to
\begin{equation}
    H_{LV} = \sqrt{\frac{\gamma}{\alpha t_C}} X \otimes \sigma_x\,,
\end{equation}
where $\gamma$ is the corresponding rate, $t_C$ is the coupling time and $\alpha=t_C/\Delta t$, where $\Delta t$ is the timestep we want to simulate. Subsequently we perform a measurement in the $|0\rangle,\, |1\rangle$ basis and apply $U$ if the result is 1. This way we can implement dynamics according to
\begin{align}
    \frac{\rho_S(\Delta t)-\rho_S}{\Delta t} = \gamma\left(L\rho_SL^\dagger-\frac{1}{2}\{L^\dagger L,\rho_S\}\right)\,.
\end{align}

This procedure can straightforwardly be generalised to multiple Lindblad operators.\\
Let us now consider $N$ Lindblad operators $L_k = U_k X_k$, $U_k$ unitary and $X_k$ positive, with corresponding rates $\gamma_k \geq 0$ for $k=1...N$. We define the Hamiltonians and rates
\begin{align}
    H_{LV,k} &= \sqrt{\frac{\Tilde{\gamma}_k}{\alpha t_C}} X \otimes \sigma_x\\
    \Tilde{\gamma}_k &= N\,\gamma_k
\end{align}
for $k=1...N$. Now we choose randomly a Hamiltonian and rate $H_{LV,k}, \Tilde{\gamma}_k$ from this set (meaning each pair has a probability $p_k=1/N$ to be selected). This Hamiltonian we apply according to the beforementioned  procedure to (on average) generate the dynamics
\begin{equation}
    \frac{\rho_S(\Delta t) - \rho_S}{\Delta t} = \Tilde{\gamma}_k \left(L\rho_SL^\dagger-\frac{1}{2}\{L^\dagger L,\rho_S\}\right)\,.
\end{equation}
Iterating this for all timesteps and many realisations yields the target evolution
\begin{align}
    \frac{\rho_S(\Delta t) - \rho_S}{\Delta t} &= \sum_{k=1}^N p_k \Tilde{\gamma}_k \left(L\rho_SL^\dagger-\frac{1}{2}\{L^\dagger L,\rho_S\}\right)\\
    &= \sum_{k=1}^N \gamma_k \left(L\rho_SL^\dagger-\frac{1}{2}\{L^\dagger L,\rho_S\}\right)\,.
\end{align}

\subsection{Influence Martingale Method}
In case the Lindblad operators $L_j$ themselves satisfy a completeness relation
\begin{equation}\label{eq:completeness}
\sum_j L_j^\dagger L_j = \mathbb{I}
\end{equation}
we can act directly modify the method outlined in Subsection \ref{subsec:nPos}. The extension relies on two observations:
\begin{itemize}
\item If $\sum_j L_j^\dagger L_j = \mathbb{I}$, then for $\|\psi\|=1$ we have 
\begin{equation}
\sum_j\|L_j\psi\| = \mathbb{I}
\end{equation}
\item For any set of non-positive definite functions $\Gamma_j(t)$, we can find a set of positive definite functions $\gamma_j(t)$ and a positive function $C(t)$ such that 
\begin{equation}
\Gamma_j(t) = \gamma_j(t) + m(t).
\end{equation}
\end{itemize}
With these two realisations we can realise a master equation of the form 
\begin{equation}\label{eq:genMaster}
\frac{d}{dt}\rho(t) = -i[H,\rho(t)] - \sum_k \Gamma_k(t)\left(L_k\rho(t)L_k^\dagger - \frac{1}{2}\{L_k^\dagger L_k, \rho(t) \}\right)
\end{equation}
by choosing a set of functions $\gamma_k(t)$ and $m(t)$ as outlined above and then perform the method outlined in Section \ref{subsec:nPos} to realise the master equation 
\begin{equation}
\frac{d}{dt}\rho(t) = -i[H,\rho(t)] + \sum_k \gamma_k(t)\left(L_k\rho(t)L_k^\dagger - \frac{1}{2}\{L_k^\dagger L_k, \rho(t) \}\right).
\end{equation}
However, we modify the protocol slightly be reweighing the measurement outcomes. Measurement 0 we reweigh by 
\begin{equation}
1+C(t) \gamma^2 t^2 = 1+C(t) \gamma^2 t^2 \sum_k L_k^\dagger L_k
\end{equation}
and the other measurements by
\begin{equation}
\frac{-\Gamma_j(t)}{\gamma_j(t)}
\end{equation}
which indeed gives equation \eqref{eq:genMaster}.

\paragraph*{Remark:}
even if \eqref{eq:completeness} is not satisfied, we can still make things work. We know there exists a positive number $a>0$ such that 
\begin{equation}
\sum_j L_j^\dagger L_j < a \mathbb{I}
\end{equation}
which means there exists an operator $P$ such that
\begin{equation}
\sum_j L_j^\dagger L_j +P^\dagger P = \mathbb{I}
\end{equation}
where we rescaled all the $L_j$, $P$ by $\sqrt{a}$. We can then realise the desired master equation by choosing the rate for $P$ to be  $\gamma(t)=C(t)$. Note that this means that whenever we measure $P$, the trajectory is given probability $0$.

\twocolumngrid
\bibliography{lit}
\end{document}